\documentclass[usenatbib]{mn2e}
\usepackage{graphicx}
\usepackage{epsfig}
\def\mr{\mathrm}

\def\kms{\mr{~km\ s}^{-1} }
\def\hi{H{\sc i}~}

\def\msol{\mr{M}_{\odot}}
\def\kmsmpc{\mr{\ km}\ \mr{s}^{-1}\ \mr{Mpc}^{-1}}

\def\cmc{\ \mr{cm^{-3}}}

\def\gcmc{\ \mr{g \cmc}}
\def\solar{\odot}

\title[Quantifying the Importance of Ram Pressure Stripping in a Galaxy Group at 100 Mpc]{Quantifying the Importance of Ram Pressure Stripping \\in a Galaxy Group at 100 Mpc}
\author[E. Freeland, C. Sengupta, \& J. H. Croston]{E. Freeland$^{1}$\thanks{E-mail:
freeland@physics.tamu.edu (EF); chandra@caha.es (CS), J.Croston@soton.ac.uk (JHC)}, C. Sengupta$^{2,3}$\footnotemark[1], \& J. H. Croston$^{4}$\footnotemark[1]\\
$^{1}$George P. \& Cynthia W. Mitchell Institute for Fundamental Physics and Astronomy,\\ Department of Physics and Astronomy, Texas A\&M University, College Station, TX 77843, USA\\
$^{2}$Calar-Alto Observatory, Centro Astronomico Hispano Aleman, C/Jesus Durban Remon, 2-2 04004 Almeria, Spain\\
$^{3}$Instituto de Astrofisica de Andalucia (CSIC), Glorieta de Astronomia s/n, 18008 Granada, Spain\\  
$^{4}$School of Physics and Astronomy, University of Southampton, Southampton, SO17 1SJ, UK}

\begin{document}

\date{Accepted 2010 July 17.  Received 2010 July 12; in original form 2010 May 14}

\pagerange{\pageref{firstpage}--\pageref{lastpage}} \pubyear{2010}

\maketitle
\begin{abstract}
We examine two members of the NGC 4065 group of galaxies: a bent-double (a.k.a. wide angle tail) radio source and an \hi deficient spiral galaxy.  Models of the X-ray emitting intragroup gas and the bent-double radio source, NGC 4061, are used to probe the density of intergalactic gas in this group.  \hi observations reveal an asymmetric, truncated distribution of \hi in spiral galaxy, UGC 07049, and the accompanying radio continuum emission reveals strong star formation.  We examine the effectiveness of ram pressure stripping as a gas removal mechanism and find that it alone cannot account for the \hi deficiency that is observed in UGC 07049 unless this galaxy has passed through the core of the group with a velocity of $\sim 800 \kms$.  A combination of tidal and ram pressure stripping are necessary to produce the \hi deficiency and asymmetry in this galaxy.
\end{abstract}

\label{firstpage}


\section{Introduction}

Galaxy characteristics, like morphology and star-formation rate (SFR), are observed to change with environment where high density regions like clusters are characterized by a larger fraction of elliptical galaxies and a lower star-formation rate than seen in the field \citep{1980ApJ...236..351D,2003MNRAS.346..601G,2003ApJ...584..210G}.  These observations are known as the morphology-density and SFR-density relations, although Whitmore et al. (1991,1993) argue that these reflect a tighter and more fundamental morphology-radius relation where the distance from the cluster centre is the independent parameter.  The morphology-density relation shows distinct behavior over three separate regimes which can be characterized in terms of the projected galaxy density or the radial distance from the cluster centre (in terms of the virial radius, $R_{vir}$, defined by the \citet{1998ApJ...505...74G} as $R_{vir}\simeq 0.002\sigma_{r}\ h^{-1}_{100}$ Mpc).  In the lowest density regions farthest from cluster centres (with projected densities $< 1\ \mathrm{Mpc}^{-2}$ or radius $> 1\ R_{vir}$) the relation is flat, suggesting that the physical mechanisms responsible for changes in morphology are not effective in this regime.  At a characteristic radius of $\sim 1\ R_{vir}$ (projected densities of $1-6\ \mathrm{Mpc}^{-2}$) the fraction of intermediate type (S0) galaxies begins to increase while the late-type disk (Sc) galaxy fraction decreases and the SFR decreases sharply \citep{2003ApJ...584..210G}.  These trends continue till $\sim 0.3\ R_{vir}$ (projected density $> 6\ \mathrm{Mpc}^{-2}$) where the fraction of intermediate types decreases and the elliptical fraction dramatically increases \citep{2003MNRAS.346..601G}.  The behavior of these relations over the separate regimes indicates a change in the dominant physical mechanisms influencing the evolution of galaxies across these environments.  

There are a variety of physical mechanisms that could be responsible for altering galaxy morphology in dense environments, including ram pressure stripping, tidal interactions, galaxy harassment, strangulation, and major and minor mergers.  The vast majority of galaxies in the universe reside in groups which are small dynamical systems typically containing a handful of large ($\sim L_*$) galaxies and a large number of smaller galaxies \citep{1983ApJS...52...61G, 1987ApJ...321..280T,2004MNRAS.348..866E,2007ApJ...671..153Y,2008A&A...479..927T}.  These systems have velocity dispersions between $\sim 30-500 \kms$ and intragroup medium (IGM) densities that are not well constrained.  Thus, it has been thought that tidal interactions are likely to dominate in this environment while ram pressure stripping and strangulation are less likely to be important.  

Galaxy groups, according to the hierarchical scenario of the formation of large-scale structure, are the building blocks of rich clusters of galaxies.  Since most galaxies exist in groups these are important sites in which to investigate the physical mechanisms responsible for the observed morphology and SFR-density relations.  Two important questions arise:
\begin{itemize}
\item To what extent are field galaxies and intergalactic gas preprocessed by the group environment before they are incorporated into galaxy clusters?
\item Which physical mechanisms are important in altering the morphology and SFR of galaxies in the group environment?
\end{itemize}
Here we present observations of the NGC 4065 group, located in the Coma Supercluster.  This group is a unique laboratory for studying ram pressure stripping as we have two independent means of measuring the density of the intergalactic gas in this system which also contains an \hi deficient, edge-on, Sc galaxy (UGC 07049).  This group has an average velocity of $6995 \pm 48 \kms$ ($z=0.0233$), a velocity dispersion of $416\pm35 \kms$, and extended X-ray emission from the intragroup medium \citep{2004ApJ...607..202M}.  Our VLA D-array data show numerous interactions among HI rich group members outside of the core of the group  \citep{2009AJ....138..295F}.  \citet{2006A&A...449..929G} classify UGC 07049 as strongly HI deficient using Arecibo data and find it has a total HI mass of $2\times 10^9\ \msol$.  This group is also known as RASSCALS NRGb 177 \citep{2000ApJ...534..114M}, GEMS NGC 4065 \citep{2004MNRAS.350.1511O}, and GH 98 \citep{1983ApJS...52...61G}.  

At the distance of this group $1 \arcmin$ corresponds to $\sim 30$ kpc.  We use a Hubble constant of $75 \kmsmpc$.

\section{Observations}

\subsection{GMRT Data}
The NGC 4065 group was observed with the Giant Metrewave Radio Telescope (GMRT) at 610 MHz in the standard continuum observing mode in August of 2007 for 8 hours including calibration.  Both the upper and lower sidebands of the correlator were used with 16 MHz bandwidth in each.  Here we present only upper sideband data.

The group was observed in H{\sc i} 21 cm line using the GMRT in May of 2008.  At $1420$ MHz, the system temperature and the gain (K/Jy) of the instrument are 76K and 0.22 respectively. The observations were carried out in the Indian polar mode. The baseband bandwidth used was 16 MHz, giving a velocity resolution of 27 km s$^{-1}$. The on source integration time was 15 hours.  The pointing centre for the  H{\sc i} observations was $12^{\rm h}\ 04^{\rm m}\ 01.^{\rm s}5\ +20^\circ\ 13^\prime\ 54.34\arcsec$ in J2000 co-ordinates.

The radio data were reduced using {\tt AIPS} (Astronomical Image Processing System) using standard procedures. Bad data due to dead antennas and radio frequency interference (RFI) 
were flagged and the data were calibrated for amplitude and phase using the primary and secondary calibrators. The primary calibrator was also used as the bandpass calibrator. The 20 cm radio continuum maps were made using the self calibrated line free channels of the observations. The radio continuum  was then subtracted from the data using the {\tt AIPS} tasks {\tt UVSUB} and {\tt UVLIN}. The final 3-dimensional deconvolved H{\sc i} data cubes were then obtained from the continuum subtracted data using the task {\tt IMAGR}. From these cubes the 
total H{\sc i} images and the H{\sc i} velocity fields were extracted using the task {\tt MOMNT}.

\subsection{XMM-Newton Data}
A short archival {\it XMM-Newton} observation of the environment of radio source NGC\,4061
(observation ID 0112271101, carried out on 2003 June 30th) was used to study
the X-ray emission from the intragroup medium. The X-ray data were reprocessed
using the {\it XMM-Newton} SAS version 6.0.0, and the latest calibration files
from the {\it XMM-Newton} website. The pn data were filtered to include only
single and double events (PATTERN $\leq 4$), and FLAG==0, and the MOS data were
filtered according to the standard flag and pattern masks (PATTERN $\leq 12$
and \#XMMEA\_EM, excluding bad columns and rows). The dataset was affected by
background flares, so filtering for good time intervals was applied to ensure
that accurate measurements could be made in low surface-brightness regions.
Lightcurves in the 10 -- 12 keV (MOS) or 12 -- 14 keV (pn) energy bands were
used to identify time periods of high background rate, and thresholds of 0.4
cts s$^{-1}$ (MOS) and 0.7 cts s$^{-1}$ (pn) were applied. The exposure times
remaining after GTI filtering were 5677 s, 6032 s and 3335 s for the MOS1, MOS2
and pn cameras, respectively.

Spectral and spatial analysis were carried out using the filtered {\it
XMM-Newton} events lists. The background was accounted for accurately using the
double subtraction method described in \citet{2008MNRAS.386.1709C}, which makes
use of filter-wheel closed datasets to constrain the instrumental and particle
components of the background. Both source and background (filter-wheel closed)
events lists were vignetting corrected using the {\sc sas} task {\it
evigweight}. The background datasets were scaled to account for differences in
the level of the instrumental and particle background components using a
weighting factor consisting of the ratio of source to background 10 -- 12 keV
count rate in a large background annulus. For surface brightness profile
analysis, a background profile was first obtained by extracting a profile
matched to the source profile from the weighted background file. The difference
in background level between source and background profiles in the outer regions
(where source emission should be negligible) was used as a second background
component to subtract off the soft X-ray background from Galactic and cosmic
X-ray components. For spectral analysis an equivalent process was used: spectra
were obtained from the source and appropriate background events lists for both
target and local background extraction regions. The Galactic/cosmic X-ray
background contribution was modeled by fitting an X-ray background model to
the local background spectrum consisting of two {\it mekal} models to account
for emission from the Galactic bubble and a power-law model absorbed by the
Galactic $N_{H}$ in the direction of the target to account for the cosmic X-ray
background. The {\it mekal} temperatures were allowed to vary, but the
power-law index was fixed at $\Gamma = 1.41$ \citep{2002A&A...389...93L}. The
normalizations of all three components were allowed to vary. For each source
spectrum, we used the corresponding filter-wheel closed spectrum as a
background spectrum, to account for particle background, and a fixed X-ray
background model consisting of the best-fitting model from the fit to the
outer, source-free region, with the normalizations of each component fixed at
the best-fitting values scaled to the appropriate area for the source
extraction region.

We extracted spectra for the group emission from an annular region of
inner radius 2 arcmin (to exclude any AGN-related X-ray emission) and
outer radius 6.7 arcmin. The spectra were fitted with a {\it mekal}
model in additional to the fixed X-ray background model as described
above. The best-fitting model parameters for a joint fit to the MOS1
and pn spectra in the energy range 0.3 -- 7.0 keV were $kT =
1.31^{+0.14}_{-0.23}$ keV and $Z = 0.33^{+0.28}_{-0.19}$ Z$_{\solar}$
using a Galactic $N_{\rm H} = 2.3 \times 10^{20}$ cm$^{-2}$, giving
$\chi^{2} = 234$ for 186 d.o.f.

\begin{figure}     
\includegraphics[width=84mm]{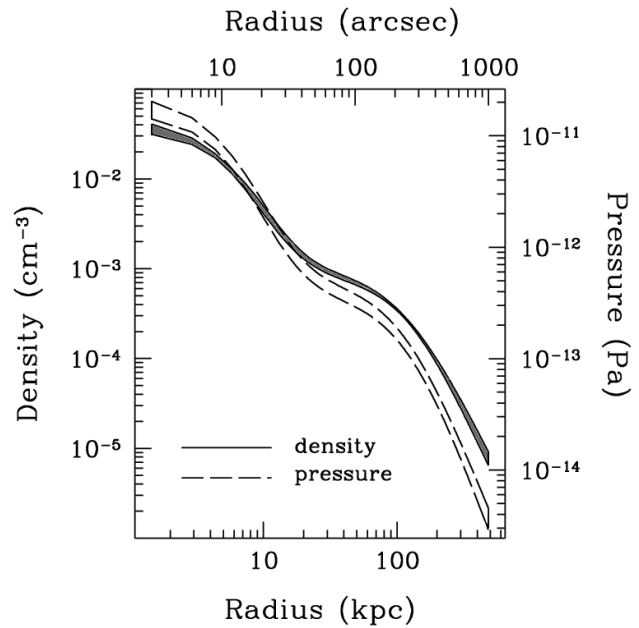}
\caption{ Number density and pressure profiles from the model fits to the XMM-Newton data showing the X-ray emitting gas in this group.  The pressure profile is calculated from the density profile assuming temperature, $kT= 1.31$ keV.  Uncertainties (1$\sigma$) in the density and pressure are illustrated by the width of the curves.}\label{fig:xrayprof}
\end{figure}

Surface brightness profiles were extracted from the MOS1 and pn events
lists, which showed evidence for both an inner galaxy-scale halo and a
flatter, group-scale component. A single $\beta$-model fit to the profiles
was therefore a poor fit. We fitted the profiles with the
projected double-beta model as described in \citet{2008MNRAS.386.1709C}, which is
the surface brightness profile corresponding to a gas density profile of:
\small
\begin{equation} n(r) = n_0 \left[ \left(1 +
\frac{r^{2}}{r_{c,in}^{2}}\right)^{-3\beta_{in}/2} + N \left(1 +
\frac{r^{2}}{r_{c}^{2}}\right)^{-3\beta / 2} \right] \end{equation}
\normalsize
where $N$ is the relative normalization of the two $\beta$ model component.
Before fitting, each model was convolved with the {\it XMM-Newton} PSF based on
the on-axis parameterization described in the {\it XMM-Newton} CCF files
XRT1\_XPSF\_0006.CCF, XRT2\_XPSF\_0007.CCF and XRT3\_XPSF\_0007.CCF. The
Markov-Chain Monte Carlo (MCMC) method for exploring the 6-dimensional
parameter space for this model described in \citet{2008MNRAS.386.1709C} was used to
determined the best fitting model parameters for the gas density profile, with
the joint $\chi^{2}$ value for the three profiles as the likelihood estimator.
Plausible ranges for each parameter were estimated by carrying out extreme fits
and these were used as priors for the MCMC method.  Quantities derived from the
surface brightness model fits, including X-ray luminosity and pressure at a
given radius, are obtained by determining that quantity for each model fit and
then obtaining the Bayesian estimate for the quantity in question.
Uncertainties on the derived quantities are the minimal 1-dimensional interval
enclosing 68 per cent of the values for the given quantity, and so correspond
to the $1\sigma$ intervals for one interesting parameter. Some of the
parameters are strongly correlated, leading to large uncertainties on the
individual model parameters; however, tight constraints can be obtained on
derived quantities such as pressure and luminosity. The best-fitting model
parameters were $\beta_{in} = 1.2$, $r_{c,in} = 13.1$ arcsec, $\beta = 1.2$,
$r_{c} = 256$ arcsec, and $N = 0.02453$ with $\chi^{2} = 24.4$ for 16 d.o.f.  Figure \ref{fig:xrayprof} shows the number density profile for the hot intergalactic gas in the core of this group.

\section{Group Dynamics}

We rely on the redshift survey of this group performed by \citet{2004ApJ...607..202M} which finds 74 members with an average velocity of $6995 \pm 48 \kms$ and a velocity dispersion of $416 \pm 35 \kms$.  The group has two compact subgroups, UZG-CG 156 (including NGC 4076) and UZC-CG 157 (including NGC 4061), which were identified by the 3D Updated Zwicky Catalog \citep{1999PASP..111..438F} solely on the criterion of compactness \citep{2002A&A...391...35F}.  These two subgroups are not distinguishable by velocity.  However, there does appear to be an extended tail in the distribution toward lower velocities.  The galaxies in this extended tail are clustered on the sky near UZC-CG 156 and may represent a filament toward nearby ($\sim 6^{\circ}$ away) rich cluster Abell 1367 whose velocity is $6595 \kms$.

If we assume that the extended tail is not significantly dynamically associated with the NGC 4065 group, and we remove those velocities from the sample, then the velocity dispersion becomes $220\pm30 \kms$.  The errors on this velocity dispersion are determined by generating bootstrap samples.

The lack of simple gaussianity in the distribution of velocities and the strong spatial bimodality in the diffuse X-ray emission indicates that the system is not in dynamical equilibrium.  If we assume the that group potential is deepest at the peak of the velocity histogram then the X-ray gas of this peak is located at a velocity of $\sim 7150 \kms$.  The velocity of NGC 4065 ($v=6326 \kms$), the early-type galaxy immediately east of radio galaxy NGC 4061, puts it likely outside of the core of the group despite its projected appearance.  The velocity of bent-double radio galaxy, NGC 4061, as measured by \citet{2004ApJ...607..202M} is $7393 \pm 32 \kms$.  Previous radial velocity measurements have given this source a velocity of $7203 \pm 27 \kms$ \citep{1991trcb.book.....D} and $7369 \pm 52 \kms$ \citep{1999PASP..111..438F}.     

\begin{figure}     
\includegraphics[width=84mm]{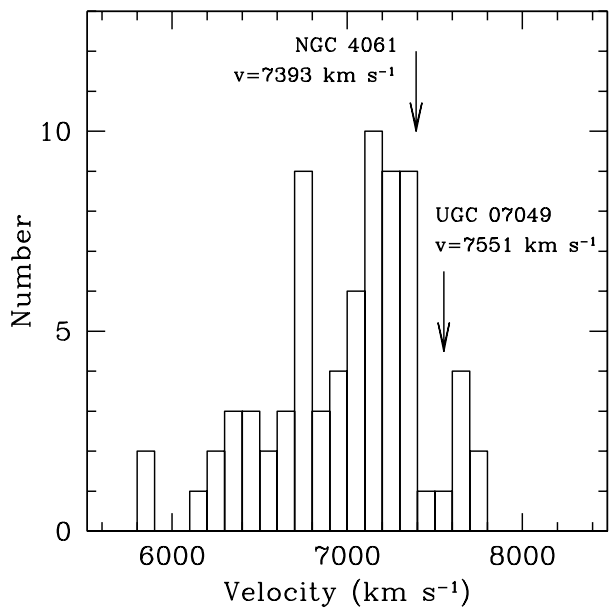}
\caption{Histogram of velocities for the NGC 4065 group of galaxies from the data in \citet{2004ApJ...607..202M}.  The velocity of bent-double radio source NGC 4061 and that of spiral galaxy UGC 07049 are indicated with arrows.  The bins are $100 \kms$ wide.} \label{fig:hist}
\end{figure}

\begin{figure*}     
\includegraphics[width=168mm]{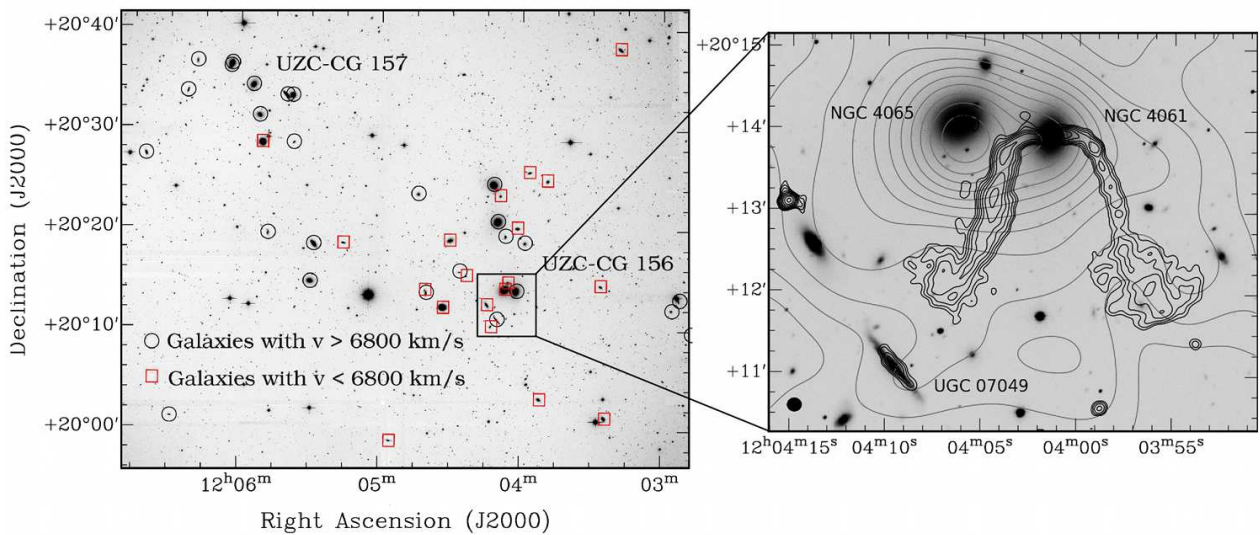}
\caption{ {\it Left:} The spatial locations of galaxies with recessional velocities greater than $6800 \kms$ are marked with black circles.  The spatial locations of galaxies with recessional velocities less than $6800 \kms$ are marked with red squares.  The underlying image was taken in the optical R band.  There are approximately 20 more galaxies identified as group members by \citet{2004ApJ...607..202M} which are beyond the edge of the optical image, mainly to the north and west.  {\it Right:} XMM X-ray (grey) and GMRT 60cm radio continuum (black) contours overlaid on an optical R band image of the core of the NGC 4065 galaxy group, also the core of compact subgroup UZC-CG 156.  Radio continuum emission from the disk of spiral galaxy UGC 07049 indicates it has a high star formation rate.  The 60cm radio contours start at $1.4$ mJy $\mathrm{beam}^{-1}$ and increase by $\sqrt{2}$.  The $7.8\arcsec\times6.7\arcsec$ 60cm beam is shown in the lower left corner.  At the distance of this group $1^{\prime}$ corresponds to 30 kpc. } \label{fig:lgt6800}
\end{figure*}

\section{Bent-Double Radio Source NGC 4061}

We follow the method for using bent-double radio sources as probes of
intergalactic gas outlined in \citet{2008ApJ...685..858F} with a few additional
caveats.  We are assuming that the jet bulk flows have decelerated to non-relativistic speeds on kilo-parsec scales, and certainly by the time they are bent by the motion of the host galaxy ($\sim 15$ kpc from the core of the radio galaxy).  This assumption is borne out by the symmetric fluxes in the two jets (see Figure \ref{fig:lgt6800}), were they still relativistic then beaming effects would produce asymmetric fluxes which would be easily observable in these data.  We also consider additional pressure in the jets from entrained material.

Bent-double radio sources can be used to measure the density of intergalactic
gas by the application of Euler's equation
\citep{1979Natur.279..770B,1979ApJ...234..818J,1980AJ.....85..204B}.  The
time-independent Euler equation describes the balance of internal and external
pressure gradients, 

\begin{equation} \frac{\rho_{\mbox{\tiny IGM}}
v_{gal}^2}{h}=\frac{\rho_{j}v_{j}^2}{R} \end{equation}  

where $\rho_{\mbox{\tiny IGM}} v_{gal}^2$ is the external ram pressure felt by
the radio galaxy as it travels through the IGM, $\rho_{j}v_{j}^2$ is the
internal pressure of the jet, $h$ is the width of the jet, and $R$ is the
radius of curvature of the jet.  We estimate the speed of the radio galaxy
using the group velocity dispersion.  The pressure in the jets is determined at
a position in the jet immediately before it bends using the minimum synchrotron
pressure as outlined in \citet{1987ApJ...316...95O}.  
Standard equipartition of energy between relativistic particles and magnetic
fields is assumed.  We assume a radio spectral index of $\alpha =0.55$ in the
jets through the point where they bend as is seen in other FRI sources with
high-resolution, multi-frequency radio data
\citep{2008ASPC..386..110L,2005ApJ...626..748Y}.  Our values for the minimum
synchrotron pressure in the jets are in good agreement with other measurements
for similar radio sources \citep{1994ApJ...436...67V,2000ApJ...530..719W}.  

Observations of straight FRI sources show that the minimum synchrotron pressure in the jets does not balance the observed external pressure from the intergalactic gas as measured by modeling its X-ray emission \citep{1984ApJ...286...68B,2002MNRAS.336.1161L,2008MNRAS.386.1709C}.  The standard assumption is that entrained thermal protons are likely responsible for the additional energy density above that provided by the relativistic particles and magnetic fields in the jets.  This entrained material must have a lower temperature than the surrounding hot gas because the X-ray surface brightness decreases at the locations of the radio lobes \citep[e.g.][]{2003MNRAS.346.1041C}.  Here we consider both the case without entrained material and the case where entrained material provides five times the energy density of the relativistic electrons.  The ratio of energy density between the relativistic electrons and the entrained material is not well constrained, however, a factor of five is typical on scales of $15-20$ kpc (Croston \& Hardcastle 2010 {\it in prep}).

We find an intergalactic gas density of $2 \pm 0.5 \times 10^{-27} \gcmc$ or $2 \pm 0.5 \times 10^{-3} \cmc$ when entrained thermal protons provide additional internal pressure in the jets.  With the standard equipartition minimum synchrotron pressure from relativistic electrons only, the density of the intergalactic gas that NGC 4061 is traveling through is $4 \pm 1 \times 10^{-28} \gcmc$ or $4 \pm 1 \times 10^{-4} \cmc$.  This intergalactic gas density, for the case including entrained protons, is very similar to the density in the model fit to the X-ray emitting intergalactic gas at the projected position of NGC 4061 ($\sim 30$ kpc from the X-ray centre).  We have good reason to believe that NGC 4061 is located within the X-ray emitting gas because the X-ray contours trace the flaring ends of the jets very closely.

\section{Star-Forming Galaxy UGC 07049}

\vspace*{0.05in}

UGC 07049 is a late type spiral at an angular distance of 3.4$^{\prime}$ from
NGC4061. It has a radial velocity of 7551 km s$^{-1}$ and is of the
morphological type Sc. As has been mentioned earlier, this galaxy shows significant \hi deficiency.  Additionally, it may be traveling through the X-ray emitting hot gas in the core of this group which makes it an interesting target in which to study how a hot IGM affects the neutral gas content of spiral galaxies.

\begin{figure}     
\includegraphics[width=84mm]{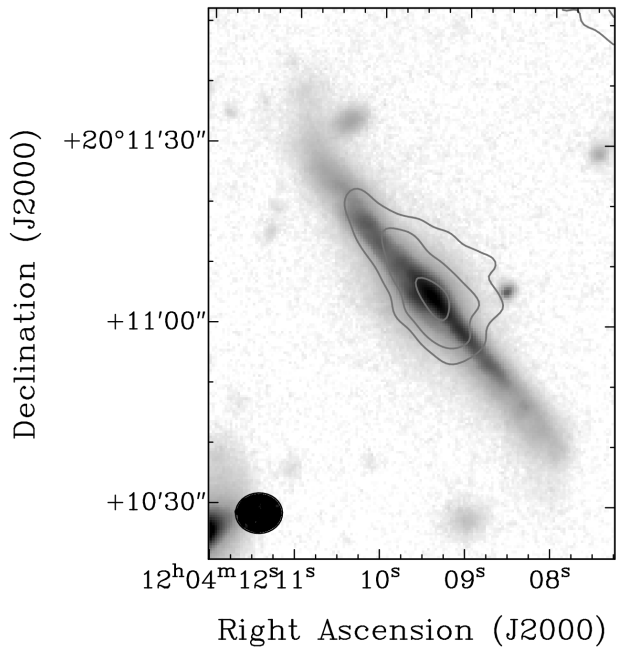}
\caption{$20$ cm radio continuum emission from UGC 07049 shown as contours on an optical R band image. The lowest contour is $0.8$ mJy $\mathrm{beam}^{-1}$ and contours increase by $\sqrt(2)$.  The beam is shown in the lower left corner of the image.  This emission is the result of star formation in the disk of the galaxy.}
    \label{fig:20cont}
    \end{figure}

\subsection{ H{\sc i} Content and Distribution}

\begin{figure}
\includegraphics[width=84mm]{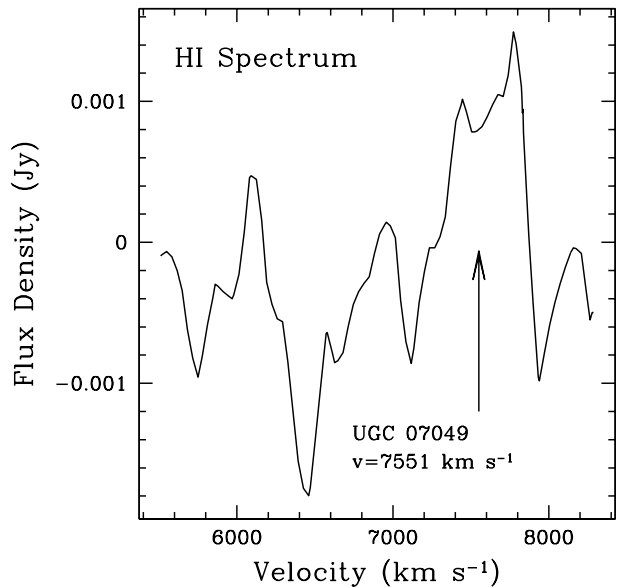}
\caption{Interferometric H{\sc i} spectrum of UGC 07049 from our GMRT data.}
\label{fig:spec}
\end{figure}

\begin{figure}      
\includegraphics[width=84mm]{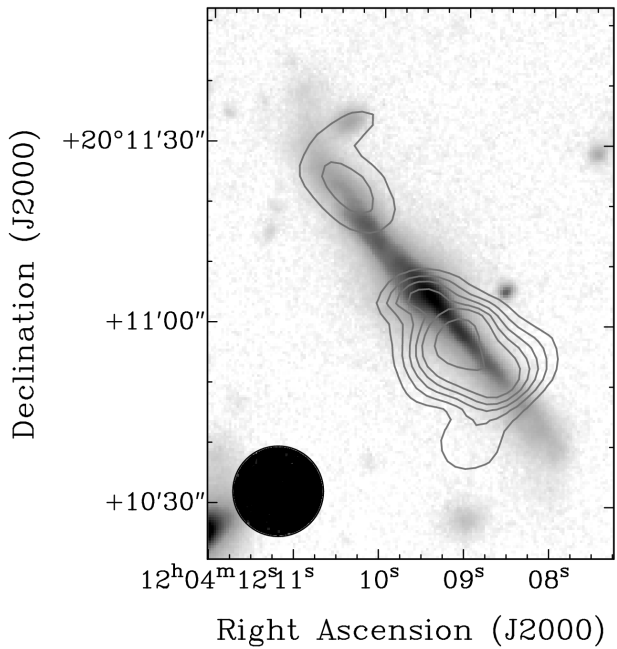} 
\caption{High resolution ($15\arcsec$) total  H{\sc i} contours for UGC7049 overlaid on an optical R band image. The  H{\sc i} column density contour levels are $8.3\times10^{19} \times$ (3, 5, 7, 9, 11, 15, 20).}
    \label{fig:hires}
    \end{figure}

Figure \ref{fig:hires} shows a high resolution ($15 \arcsec \times 15\arcsec$) H{\sc i} column density image of UGC 07049 overlaid on an optical R band image. The H{\sc i} distribution in the disk is asymmetric, with a peak in the \hi south of the centre of the galaxy.  A distinct feature is that the H{\sc i} disk is similar in extent to the optical disk. In normal late type spirals of similar morphological type, typical H{\sc i} disks are found to be 1.5 to 2 times larger in extent than the optical disk \citep{1997A&A...324..877B}.  Figure \ref{fig:spec} shows the interferometric H{\sc i} spectrum of UGC 07049 from our GMRT data. The integrated flux density as estimated from this spectrum is 0.86 Jy km s$^{-1}$, which matches well with the single dish value reported in \citet{1989gcho.book.....H}($0.85$ Jy km s$^{-1}$) and is slightly lower than that reported from Arecibo observations by \citet{2005ApJS..160..149S}($1.12$ Jy km s$^{-1}$).  The 0.86 Jy km s$^{-1}$ integrated flux density corresponds to an \hi mass of $2 \times 10^9\ \msol$. A reasonable match of the integrated line flux densities of GMRT data and single dish observations imply that we have not lost any diffuse extended emission and the H{\sc i} disk is indeed similar in size as the optical disk. 

The parameter that indicates whether a galaxy has lost gas compared to a field galaxy of similar size and similar morphological type is typically known as the `H{\sc i} deficiency' and is given by
\begin{equation}
{\it def_{\rm {H_{I}}}=log{{\frac{M_{H_{I}}}{D_{l}^{2}}}|_{predicted}}~-~log{{\frac{M_{H_{I}}}{D_{l}^{2}}}|_{observed}}} 
\end{equation} 

where $M_{H_{I}}$ is the total H{\sc i} mass of a galaxy and $D_{l}$ is the optical major isophotal diameter (in kpc) measured at or reduced to a surface brightness level m$_B$ = 25.0 mag/arcsec$^{2}$.

The `predicted' field galaxy value of H{\sc i} surface density for morphological type Sc has been taken from \cite{1984AJ.....89..758H}. While \cite{1984AJ.....89..758H} used the UGC blue major diameters for $D_{l}$, in this work RC3 major diameter have been used. To take care of the difference in the surface matter density that result from the use of RC3 diameters, a value of 0.08 \citep{1991trcb.book.....D} has been added to the predicted surface density given by \cite{1984AJ.....89..758H}.  Assuming a distance of 100.7 Mpc, derived using the optical velocity of UGC 07049 and using its RC3 diameter value 0.91$^{\prime}$, the H{\sc i} deficiency of the galaxy is found to be $\sim$ 0.41. This implies that the galaxy is $\sim$ 2.6 times deficient in H{\sc i}.

\subsection{Ram Pressure Stripping}

Possible causes for the \hi deficiency include tidal interactions, ram pressure stripping, or a combination of these two processes. In the H{\sc i} images we do not see any H{\sc i} tidal extensions, although the stellar disk appears to be warped on the north-east end.  Recent studies have shown that in groups with X-ray bright IGM ram pressure alone or tidally aided ram pressure is capable of removing a considerable fraction of the gas from constituent galaxies \citep{2007MNRAS.378..137S,2006MNRAS.369..360S,2006MNRAS.370..453R,1997AJ....114..613D}.  The \hi data presented here show the gaseous disk of UGC 07049 to be of similar size to the optical disk, whereas normal galaxies are known to have larger H{\sc i} disks compared to their optical disks.  Truncated HI disks are common in clusters of galaxies and in many cases are thought to result from ram pressure stripping. The low column density gas from the outer edges of the disk is swept out by the ram pressure offered by the dense intracluster gas, leaving behind a reduced H{\sc i} disk.

The NGC 4065 group has been detected to have a hot IGM and also a bent radio jet indicating resistance from the IGM. The spiral UGC 07049 is located within the projected X-ray emitting gas. Using some simple calculations we will explore whether ram pressure in this group is strong enough to strip gas from UGC 07049. We have followed the method outlined in \citet{2007MNRAS.378..137S} for calculating the possible gas loss from UGC 07049 due to ram pressure. 

Ram pressure stripping will be effective for a galaxy when the H{\sc i} surface
density, $\sigma_{\mu}$, is less than $\rho_{\mbox{\tiny IGM}}{v}_{gal}^{2}/(2\pi~G\sigma_{*})$, where
$\sigma_{*}$ is the stellar surface density, $\rho_{\mbox{\tiny IGM}}$ is the local IGM
density and $v_{gal}$ is the velocity with which the galaxy is moving through the medium.  We use the local projected IGM density of 4$\times$10$^{-4} \cmc$ from the X-ray gas profile, and for the velocity of UGC 07049 we consider both the $220 \kms$ and $416 \kms$ velocity dispersion.  The stellar surface
density for UGC 07049 is estimated from its 2MASS K-band magnitude and K-J color.
Relating the mass to light ratio in the K band ($M/L_{K}$) to the K-J
colors \citep{2001ApJ...550..212B} along with the relation of $L_{K}$ to
the absolute magnitude in K-band ($M_{K}$) \citep{1994ApJS...95..107W}, we
estimate the stellar surface density to be 0.0194 gm cm$^{-2}$.  With this value
for $\sigma_{*}$, the critical H{\sc i} surface density
$\sigma_{\mu}$ beyond which ram pressure can strip gas off the galaxy, is
estimated to be $3.8 \times 10^{-5}\ \mathrm{g cm}^{-2}$ or 2.3$\times$10$^{19}$ cm$^{-2}$.  We assume the H{\sc i}
surface density distribution inside UGC 07049 to be of constant thickness and with a Gaussian
profile \citep{1980A&A....83...38C},  

\begin{equation} \sigma(r)= \sigma_{0}~2^{-{r^{2}}/{r_{H}^{2}}} \end{equation}

 where $r_{H}$ is the radius within which half the  H{\sc i} mass is present.
For $\sigma_{0}$ an average value of 20 M$_\odot$ pc$^{-2}$, as seen in normal late type spirals \citep{2005JApA...26...34O} in group environments was used. The central H{\sc i} column density of UGC 07049 also could have been used but the disk was too disturbed and there was a central HI depletion in the high resolution map and thus the map was not used for estimating the $\sigma_{0}$ value.  Integrating the surface density distribution to equal the entire observed \hi mass allows us to solve for $r_{H}$ and find a value of $5$ kpc.  The radius corresponding to $\sigma_{\mu}$ is then $10$ kpc (for the case where $v_{gal}=416 \kms$).  Finally the H{\sc i} mass outside this radius, which would be able to be ram pressure stripped, was estimated to be 7$\times$10$^{7}$ M$_\odot$. A similar estimate of the stripped H{\sc i} mass was done with the $220 \kms$ velocity dispersion and, understandably, produced a smaller mass loss.  These calculations assume the galaxy is traveling face-on through the IGM and they do not take into account any of the galaxy's orbital history.  

Compared to the observed H{\sc i} deficiency the amount of ram pressure stripped \hi is very small, indicating that under the assumed criteria ram pressure alone would not contribute to gas loss in a significant way. However, tidal interactions have been observed to remove large amounts of \hi gas from galaxies in groups \citep{2009AJ....138..295F} and can increase the effectiveness of ram pressure stripping.  In this scenario, tidal interactions may disturb the gas disk, pull gas out, and reduce the column density enabling even weak ram pressure to strip off low column density gas in large quantities. It is quite possible that UGC 07049 has undergone such stripping in this group given the richness of the system, a hostile X-ray emitting dense IGM, lower than normal gas content, a truncated and asymmetric H{\sc i} disk, and the disturbed stellar disk. In addition to these, the possibility of this galaxy to have passed through the group core and thereby experience a higher IGM density cannot be ruled out.  A trip through the core of this group, experiencing intergalactic gas densities nearly ten times higher, can ram pressure strip enough \hi gas to explain the deficiency only if the galaxy velocity is as high as $\sim 800 \kms$.

\subsection{Star Formation Rate}
UGC 07049 has a $1420$ MHz radio continuum flux density of 10.9 mJy.  At a distance of 100.7 Mpc this corresponds to an upper limit on the star formation rate of $7\ \msol\ \mr{yr}^{-1}$ using the simple relation in \citet{2001ApJ...554..803Y}.  This assumes a Salpeter initial mass function integrated over stars with masses ranging from $0.1 \msol$ to $100 \msol$.

With a measured \hi mass of $2\times 10^9\ \msol$ and a star formation rate of $7\ \msol\ \mr{yr}^{-1}$, UGC 07049 will exhaust its \hi gas in $\sim 300$ Myr.  For comparison, the crossing time for this group, assuming a diameter of 1 Mpc, is $\sim 2$ Gyr.  Thus, the \hi deficiency seen in UGC 07049 may also be a result of an elevated star formation rate.    

\section{Summary}

We examine two objects in the NGC 4065 group of galaxies: a bent-double radio source and an \hi deficient spiral galaxy.  We use X-ray observations and the bent-double radio source to probe the density of intergalactic gas in this group.  For the \hi deficient spiral, UGC 07049, we calculate the effectiveness of ram pressure stripping as a gas removal mechanism and find that it alone is not strong enough to produce the amount of deficiency that is observed.  An elevated star-formation rate is also observed in this galaxy.  A combination of tidal and ram pressure stripping, with help from the elevated star-formation rate, are likely strong enough to produce the observed \hi deficiency.

\section*{Acknowledgments}
We thank the staff of the GMRT who have made these observations possible. GMRT is run by the National Centre for Radio Astrophysics of the Tata Institute of Fundamental Research.  This research has made use of the NASA/IPAC extragalactic database (NED) which is operated by the Jet Propulsion Laboratory, Caltech, under contract with the National Aeronautics and Space Administration.

\bsp
\label{lastpage}
 
\end{document}